\documentclass[11pt,a4paper]{article} 
\usepackage{amsfonts}

\usepackage[dvips]{graphicx}
\usepackage{amsthm}
\usepackage{amsmath}
\usepackage{epsfig}
\usepackage{t1enc}
\usepackage{amssymb}
\usepackage{latexsym}
\usepackage{bm}
\usepackage[hang]{caption}

\def\sp{{(S)}} \def\H{{\cal H}} \def\K{{\cal K}}
\def\z{\zeta}  
\def\Pz{\Phi^{(0)}} \def\tA{\tilde A} \def\te{\tilde \eta}
 \def\xs{\xi_*} 
\def\L{\Lambda} \def\sm{s_{min}} \def\lr
{\l \mbox{\hskip -1.46mm\raisebox{0.6mm}{\Large -}}}
\def\I{{II}} 
\def\r{\rho} \def\d{\delta} \def\D{\Delta} 
\def\s{\sigma} \def\l{\lambda}

\def\E{{\cal E}}  \def\B{{\cal B}} 
\def\a{\alpha} \def\b{\beta}

\def\g{\gamma} \def\G{\Gamma}
 \def\Q{{\cal Q}} \def\R{{\cal R}}
 \def\tQ{\tilde \Q} 
  
\def\hX{\hat X} \def\k{\kappa} \def\tB{\tilde \B}
 \def\X{{\cal X}} 
 \def\0{{(0)}} 
\def\o{\omega} \def\O{\Omega} \def\th{\theta}
\def\e{\epsilon} 
\def\Bl{\Big(} \def\Br{\Big)} \def\BL{\Big[} \def\BR{\Big]}
\def\tH{{\tilde \H}} \def\Pid{\Phi_{id}} \def\tx{{\tilde \xi}}
\def\Y{{\cal Y}} \def\hcX{\hat \X}
\def\kp{k_\|} \def\u{c} \def\bs{{\bar s}}
\def\Bla{\Big\langle} \def\Bra{\Big\rangle}
\def\cG{{\cal G}} \def\dP{\D \Phi} \def\Z{{\cal Z}}

\title{\bf Effect of radiation-like solid on CMB anisotropies}

\author{Vladim\'ir Balek\footnote{e-mail
address: balek@fmph.uniba.sk}\ \ and Matej \v Skovran
\footnote{e-mail address: skovran@fmph.uniba.sk}
\\
{\it Department of Theoretical Physics, Comenius University,
Bratislava, Slovakia}}

\begin{document}

\renewcommand{\figurename}{Fig.}

\maketitle \maketitle\abstract We compute the power in the lowest
multipoles of CMB anisotropies in the presence of radiation-like
solid, a hypothetical new kind of radiation with nonzero shear
modulus. { If only the ordinary Sachs-Wolfe effect is taken into
account, the shear modulus to energy density ratio must be in
absolute value of order $10^{-5}$ or less for the theory to be
consistent with observations within cosmic variance. With the
integrated Sachs-Wolfe effect switched on, the constraint is
relaxed almost by two orders of magnitude.}


\section{Introduction}

The observed acceleration of the universe is usually interpreted
as an effect of dark energy, originating in null oscillations of
quantum fields. An alternative explanation is that what
accelerates the universe is a solid with negative pressure to
energy density ratio $w$. The idea appeared shortly after the
effect was discovered \cite{{bs},{bbs}}, and the underlying theory
was extensively studied afterwards
\cite{{lei},{bm1},{bm2},{bm},{bp},{kum},{gru},{end},{bar},{sit}}.

Given the attention paid to scenarios in which the solid has
negative $w$, it is natural to ask what difference would it make
if the value of $w$ of the solid was positive. This question was
addressed in \cite{bal}. One of the two scenarios explored there
was that in a universe filled with radiation there appeared a {\it
radiation-like solid}, which had, in addition to $w = 1/3$, a
constant shear modulus to energy density ratio $\xi$. A possible
materialization of such solid could be a Coulomb crystal with
relativistic Fermi gas of moving charges; another, more
speculative possibility could be a network of `spring-like'
strings with energy inversely proportional to length. In the
latter case $\xi$ would be negative, making the vector
perturbations unstable. This, however, would not necessarily make
the theory useless, since in the course of inflation typically no
vector perturbations are created.

In order to obtain an observable effect of a solid on superhorizon
perturbations one must assume that the solid was created with flat
internal geometry and nonzero shear stress acting in it. Such
solidification cannot take place in pure radiation, whose
particles move freely and hence the concept of shear deformation
has no sense for them. It must be linked to some other kind of
particles present in the universe, distributed anisotropically
from the start of Friedmann expansion. Such component of cosmic
medium could perhaps be formed in solid inflation
\cite{{gru},{end},{bar},{sit}}.

In the paper we investigate how the radiation-like solid, if
present in our universe, would manifest itself in the large-angle
CMB anisotropies. In section \ref{sec:evol} we describe how the
superhorizon perturbations evolve in the presence of solid and
determine their magnitude at the moment of recombination; in
section \ref{sec:eff} we derive a formula for temperature
fluctuations observed at present on Earth and calculate the CMB
power spectrum for the lowest multipole moments; and in section
\ref{sec:con} we discuss the results. A system of units is used in
which $c = 16\pi G = 1$.


\section{Evolution of perturbations}
 \label{sec:evol}

\subsection{Pure solid}

Consider a flat FRWL universe filled with an isotropic elastic
matter and add a small perturbation of spacetime metric and
distribution of matter to it. Following \cite{pol}, we will use
{\it proper-time comoving gauge} for the description of
perturbations; thus, we will assume that $\phi$ (the correction to
the $(00)$-component of the metric tensor) as well as $\d \bf x$
(the shift vector of the matter) is zero. Denote the scale
parameter by $a$, the mass density and pressure of the matter by
$\r$ and $p$, the compressional modulus of the matter by $K$ and
the shear modulus of the matter by $\mu$. The condition $\phi = 0$
means that the scalar part of the metric is
\begin{equation}
ds^{\sp 2} = a^2 [d\eta^2 + 2 B_{,i} d\eta dx^i - (\d_{ij} - 2\psi
\d_{ij} - 2E_{,ij}) dx^i dx^j],
 \label{eq:ds2}
\end{equation}
and from the condition $\d {\bf x} = 0$ it follows that the scalar
part of the energy-momentum tensor is
\begin{equation}
{T_0}^0 = \r + \r_+ (3\psi + \E), \quad {T_i}^{0\sp} = \r_+
B_{,i}, \quad {T_i}^{j\sp} = - p \d_{ij} - K (3\psi + \E) \d_{ij}
- 2\mu E^T_{,ij},
 \label{eq:Tmn}
\end{equation}
where $\r_+ = \r + p$, $\E = \triangle E$ and $E^T_{,ij}$ is the
traceless part of the tensor $E_{,ij}$, $E^T_{,ij} = E_{,ij} -
\triangle E \d_{ij}/3$. In the formulas for ${T_0}^0$ and
${T_i}^{j\sp}$ we have assumed that the perturbations are {\it
adiabatic}, that is, the entropy per particle $S$ is constant
throughout the space.

The proper-time comoving gauge allows for a residual
transformation $\d \eta = a^{-1} \d t({\bf x})$, where $\d t({\bf
x})$ is the local shift of the moment at which the time count has
started. The function $\E$ is invariant under such transformation
and the functions $B$ and $\psi$ can be written as
\begin{equation}
B = \B + \chi, \quad \psi = - \H \chi,
 \label{eq:Bps}
\end{equation}
where $\B$ is invariant and $\chi$ transforms as $\chi \to \chi +
\d \eta$.

Suppose the solid has flat internal geometry and consider a
perturbation of the form of plane wave with the comoving wave
vector $\bf k$. For $\B$ and $\E$ as functions of $\eta$ we have
two coupled differential equations of first order, coming from
equations ${{T_i}^\mu}_{;\mu} = 0$ and $2G_{00} = T_{00}$,
\begin{equation}
\B' = (3c_{S0}^2 + \a - 1)\H \B + c_{S\|}^2 \E, \quad \E' = - (k^2
+ 3\a \H^2) \B - \a \H \E,
 \label{eq:dBdE}
\end{equation}
where the prime denotes differentiation with respect to $\eta$,
$\H$ is the rate of expansion, $\H = a'/a$, and the functions
$\a$, $c_{S0}^2$ (auxiliary sound speed squared) and $c_{S\|}^2$
(longitudinal sound speed squared) are defined as $\a =
(3/2)\r_+/\r$, $c_{S0}^2 = K/\r_+$ and $c_{S\|}^2 = c_{S0}^2 +
(4/3)\mu/\r_+$. These equations must be supplemented by the
dynamical equations for an unperturbed universe,
\begin{equation}
a' = \Bl\frac 16 \r a^4\Br^{1/2}, \quad \r' = - 3\H\r_+,
 \label{eq:Eeq}
\end{equation}
and an expression for $K$,
\begin{equation}
K = \r_+ \Big( \frac{dp}{d\r}\Big)_S.
 \label{eq:K}
\end{equation}
If the matter has more than one component, it holds $\r =
\displaystyle \sum \r_i$ and $K = \displaystyle \sum K_i$, and the
second equation (\ref{eq:Eeq}) as well as equation (\ref{eq:K})
hold for each component separately.

Consider a one-component universe in which the parameters $w =
p/\r$ and $\xi = \mu/\r$ are constant. By combining the two
equations (\ref{eq:Eeq}) we obtain that $a$ is a power-like
function of $\eta$, $a \propto \eta^{2/(1 + 3w)}$, and from
equation (\ref{eq:K}) we find that the parameter $\K = K/\r$ is
constant, too, $\K = ww_+$ with $w_+ = w + 1$. Equations for $\B$
and $\E$ then combine into an equation for $\B$ whose solution are
Bessel functions multiplied by a certain power of $\eta$. In case
$w = 1/3$ the equations for $\B$ and $\E$ are
\begin{equation}
\B' = 2 \eta^{-1} \B + c_{S\|}^2 \E, \quad \E' = - (k^2 + 6
\eta^{-2}) \B - 2 \eta^{-1} \E,
 \label{eq:dBdEw}
\end{equation}
where $c_{S\|}^2 = 1/3 + \xi$; and after excluding $\E$ we obtain
an equation for $\B$,
\begin{equation}
\B'' + (c_{S\|}^2 k^2 + 6\xi\eta^{-2}) \B = 0,
 \label{eq:ddB}
\end{equation}
with the solution
\begin{equation}
\B = \sqrt{z} (a_J J_n + a_Y Y_n),
 \label{eq:solB}
\end{equation}
where $z = c_{S\|}k\eta$, $n = \sqrt{1/4 - 6\xi}$ and $J_n$ and
$Y_n$ are Bessel functions of first and second kind of the
argument $z$.

When describing CMB anisotropies it is convenient to pass to {\it
Newtonian gauge}, in which the functions $B$ and $E$ are traded
for the Newtonian potential $\phi$ and the scalar part of the
shift vector $\d {\bf x}^\sp$. Denote the potentials $\phi$ and
$\psi$ and the perturbation to the mass density $\d \r$ in the
Newtonian gauge by $\Phi$, $\Psi$ and $\overline{\d \r}$. For the
functions $\Psi$ and $\overline{\d \r}$ we can use expressions
coming from the gauge transformation (equations (7.19) and (7.20)
in \cite{mukh}) and for the function $\D \Phi = \Phi - \Psi$ we
have an expression following from Einstein equations (equation
(7.40) in \cite{mukh}),
\begin{equation*}
\Psi = \psi + \H (B - E'), \quad \D \Phi = \frac 12 a^2
\tau^{(2)}, \quad \overline{\d \r} = \d \r - \r' (B - E'),
\end{equation*}
where $\tau^{(2)}$ is the longitudinal part of the perturbation to
${T_i}^{j\sp}$ and $\d \r$ is the perturbation to ${T_0}^0$. By
inserting here from equations (\ref{eq:Tmn}) and (\ref{eq:Bps}) we
find
\begin{equation}
\Psi = \H (\B - E'), \quad \D \Phi = - \mu a^2 E, \quad
\overline{\d \r} = \r_+ (3\Psi + \E),
 \label{eq:PPdr}
\end{equation}
and after inserting into the expression for $\Psi$ from the second
equation in (\ref{eq:dBdE}) and into the expression for $\D \Phi$
from the first equation in (\ref{eq:Eeq}), we arrive at
\begin{equation}
\Psi = - k^{-2}\a  \H^2 (3 \H \B + \E), \quad \dP = 6\xi k^{-2}
\H^2 \E.
 \label{eq:PP0}
\end{equation}
For $w = 1/3$ this yields
\begin{equation}
\Psi = - 2 (k\eta)^{-2} (3 \eta^{-1} \B + \E), \quad \dP = 6\xi
(k\eta)^{-2} \E,
 \label{eq:PP}
\end{equation}
and by inserting for $\E$ from the first equation in
(\ref{eq:dBdE}), we can write $\Phi$ and $\Psi$ in terms of $\B$
only,
\begin{equation}
\Psi \propto - z^{-2} \BL\frac {d\B}{dz} - (1 - 3\xi) z^{-1}
\B\BR, \quad \D \Phi \propto 3\xi z^{-2} \Bl \frac {d\B}{dz} - 2
z^{-1} \B\Br.
 \label{eq:PP1}
\end{equation}
With $\B$ in the form (\ref{eq:solB}), these are expressions for
the potentials $\Phi$ and $\Psi$ in terms of Bessel functions and
their derivatives.

An important special case are {\it superhorizon perturbations},
whose wavelength exceeds significantly the size of the sound
horizon. The radius of the sound horizon is $r_S \sim c_{S\|}
r_h$, where $r_h = a\eta$ is the radius of particle horizon; thus,
the variable $z$ can be written as $z \sim r_S/\lr$, where $\lr =
\l/(2\pi) = ak^{-1}$ is the reduced wavelength, and the condition
on the size of the superhorizon perturbations can be written as $z
\ll 1$. In this limit it holds
\begin{equation*}
J_n \propto z^n \Big(1 - \frac {z^2}{4n_+}\Big), \quad Y_n \propto
-z^{-n},
\end{equation*}
where $n_+ = n + 1$, and if we insert this into the expression for
$\Psi$, we obtain
\begin{equation}
\Psi = A_J \Big(\frac pq z^{-m} + z^{- 2 - m}\Big) - A_Y z^{- 2 -
M},
 \label{eq:asPz}
\end{equation}
where $m = 1/2 - n$, $M = 1/2 + n$ and the constants $p$ and $q$
are defined as
\begin{equation}
p = \frac 1{4n_+} (2 - m + 3\xi), \quad q = m - 3\xi.
 \label{eq:pq}
\end{equation}
For $|\xi| \ll 1$ it holds $p \doteq 1/3$ and $q \doteq 3\xi$,
therefore $|p/q| \gg 1$ and the term proportional to $z^{-m}$ can
dominate the term proportional to $z^{-2 - m}$ while $z$ is still
small.

Consider now a universe filled with an ideal fluid and suppose
that the fluid turned into a solid with the same $w$ at some
moment $\eta_s$. As before, we pick $w = 1/3$. Suppose the
solidification was {\it anisotropic}, producing a solid with flat
internal geometry, and consider perturbations that were
superhorizon at the moment $\eta_s$; in other words, suppose $z_s
\ll 1$. The first condition means that the equations
(\ref{eq:dBdE}) are valid without modifications for all $\eta
> \eta_s$ and the second condition implies that $\Psi$ can be
written in the form (\ref{eq:asPz}) in some interval of $\eta >
\eta_s$. In a fluid, the potentials $\Phi$ and $\Psi$ coincide
and, if we restrict ourselves to their non-decaying part, they are
constant for long-wave perturbations. Denote this constant by
$\Pz$. If $|\xi|$ is not too small, $|\xi| \gg z_s^2$, the first
term in the brackets in (\ref{eq:asPz}) can be neglected at the
moment $\eta_s$ and the matching conditions reduce to
\begin{equation}
\tA_J - \tA_Y = \Pz, \quad (2 + m) \tA_J - (2 + M) \tA_Y = -
\eta_s [\Psi'],
 \label{eq:match}
\end{equation}
where $\tA_J = A_J z_s^{- 2 - m}$, $\tA_Y = A_Y z_s^{- 2 - M}$ and
the square brackets denote the jump of the function at $\eta =
\eta_s$. To determine $[\Psi']$, we observe that the density
contrast $\d = \overline{\d \r}/\r$ equals $-2 \Pz$ for long-wave
perturbations in a universe filled with fluid \cite{mukh}. After
rewriting the last equation in (\ref{eq:PPdr}) as $\d = w_+ (3\Pz
+ \E)$ and putting $w = 1/3$, we find $\E = - (9/2)\Pz$ and
\begin{equation*}
\eta_s [\Psi'] = -6 (k\eta_s)^{-2} [B'] = -6 c_{S\|}^2 z_s^{-2}
\xi \E_s = 27 c_{S\|}^2 \xi z_s^{-2} \Pz.
\end{equation*}
Our assumption about the values of $\xi$ guarantees that this is
in absolute value much greater than $|\Pz|$, so that we can
neglect the right hand side of the first equation in
(\ref{eq:match}). The solution  is
\begin{equation}
\tA_J = \tA_Y = r z_s^{-2} \Pz, \quad r = \frac {27}{2n} c_{S\|}^2
\xi,
 \label{eq:AJAY}
\end{equation}
and after inserting this into (\ref{eq:asPz}) and introducing a
rescaled time $\te$ normalized to 1 at the moment $\eta_s$, $\te =
\eta/\eta_s = z/z_s$, we obtain
\begin{equation}
\Psi = P \te^{-m} + Q z_s^{-2} (\te^{- 2 - m} - \te^{- 2 - M}),
 \label{eq:asP}
\end{equation}
where $P = (p/q) r \Pz$ and $Q = r \Pz$.

\subsection{Adding the solid to the rest of matter}

Consider a universe filled with nonrelativistic matter (dust) with
density $\r_m$ and pressure $p_m = 0$ and radiation with density
$\r_r$ and pressure $p_r = \r_r/3$, and suppose that the radiation
consists of ordinary radiation (photons and neutrinos) with the
density $\r_{rI}$ and zero shear modulus, and of radiation-like
solid with the density $\r_{rII}$ and shear modulus $\mu$
proportional to $\r_{rII}$. Both ordinary radiation and
radiation-like solid have the same dependence of $p$ on $\r$ and
hence of $\rho$ on $a$; thus, the densities $\r_{rI}$ and
$\r_{rII}$ are at any given moment equal to the same fractions of
$\r_r$ and $\mu$ is proportional to $\r_r$, $\mu = \xi \r_r$ with
constant $\xi$.

The shear modulus of ordinary solids is much less than their
mass/energy density. On the other hand, one would expect the shear
modulus of the hypothetical radiation-like solid to be {\it
comparable} with mass/energy density, as is the case with random
lattices of cosmic strings or domain walls which have both
$\xi_{net}$ (the net parameter $\xi$) equal to 4/15 \cite{bccm}.
If a radiation-like solid has $|\xi_{net}| \sim 1$, the ratio of
its density to the density of ordinary radiation would be of order
$|\xi|$, unless $\xi$ would be close to $\xi_{net}$ and hence
$|\xi|$ would be close to 1. However, as we will see, $|\xi|$ must
be much less than 1 to accommodate observations; therefore if a
radiation-like solid with $|\xi_{net}| \sim 1$ was present in our
universe, its today's density would be just a tiny fraction of the
density of radiation.

For an unperturbed universe with radiation and matter it makes no
difference whether it does or does not contain a radiation-like
solid. Let us write down the solution of equations (\ref{eq:Eeq})
in this case. Denote $\z = \eta/\eta_*$ with $\eta_* =
\eta_{eq}/(\sqrt{2} - 1)$, where the index `eq' refers to the
moment when the densities $\r_r$ and $\r_m$ are equal. From the
second equation (\ref{eq:Eeq}), written separately for radiation
and matter, we have
\begin{equation*}
\r_r \propto a^{-4}, \quad \r_m \propto a^{-3},
\end{equation*}
and from the first equation (\ref{eq:Eeq}) with $\r = \r_r + \r_m$
on the right hand side we obtain
\begin{equation}
a = a_{eq} \z (2 + \z).
 \label{eq:arm}
\end{equation}

Let us now turn to the perturbed universe. For $\B$ and $\E$ we
have equations (\ref{eq:dBdE}) with the derivatives with respect
to $\z$ instead of $\eta$ and with the replacements $\B \to \tB =
\eta_*^{-1} \B$, $\H \to \tH = \eta_* \H$ and $k \to \k =
k\eta_*$; and for $\Phi$ and $\Psi$ we have equations
(\ref{eq:PP0}) with an additional replacement $\xi \to \tx = \xi
\r_r/\r$. Denote $\z_+ = 1 + \z$, $X = \z (2 + \z)$, $X_+ = 1 + X
= \z_+^2$ and $\hX_+ = 1 + (3/4)X$. From the formulas for $\r_r$
and $\r_m$ we have $\r_r/\r_m = a_{eq}/a = 1/X$, so that $\r_r/\r
= 1/X_+$, $\r_{r+}/\r_+ = 1/\hX_+$ and $\r_+/\r = (4/3)
\hX_+/X_+$. The function $\a$ equals $(3/2)\r_+/\r$ as before and
the functions $c_{S0}^2$ and $c_{S\|}^2$ now equal
$(1/3)\r_{r+}/\r_+$ (because $K = K_r = \r_{r+}/3$) and $(1/3 +
\xi) \r_{r+}/\r_+$. Finally, for the function $\tH$ we have $\tH =
a^{-1} da/d\z$ with $a \propto X$. In this way we find
\begin{equation*}
\a = \frac {2\hX_+}{X_+},\quad c_{S0}^2 = \frac {1/3} {\hX_+},
\quad c_{S\|}^2 = \frac {1/3 + \xi}{\hX_+}, \quad \tH = \frac
{2\z_+}X,
\end{equation*}
and if we insert this into equations (\ref{eq:dBdE}) and
(\ref{eq:PP0}), we arrive at
\begin{equation}
\frac {d\tB}{d\z} =  \Bl \frac 1{\hX_+} + \frac {1 + X/2}{X_+} \Br
\frac {2\z_+}X \tB + \frac {1/3 + \xi}{\hX_+} \E, \quad \frac
{d\E}{d\z} = - \Bl \k^2 + \frac {24\hX_+}{X^2}\Br \tB - \frac
{4\hX_+}{\z_+ X} \E,
 \label{eq:dBdExi}
\end{equation}
and
\begin{equation}
\Psi = - 8 \k^{-2} \frac{\hX_+}{X^2} \Bl \frac {6\z_+}X \tB + \E
\Br, \quad \dP = \frac {24 \xi}{X^2} \k^{-2} \E.
 \label{eq:PPxi}
\end{equation}

A straightforward way to compute the functions $\Phi$ and $\Psi$
would be by solving an equation of second order for $\tB$, similar
as in the theory with one-component matter. However, if we are
interested in small values of $|\xi|$ and $\k$ only, it is
preferable to pass to equations of first order for $\Psi$ and
$\E$. As we will see, the equations simplify if we factor a
certain function of $\z$ out of $\Psi$.

By differentiating $\Psi$ and using equations for $\tB$ and $\E$
we find
\begin{equation}
\frac {d\Psi}{d\z} = - \Bl 1 + \frac {2\hX_+}{X_+} \Br \frac
{2\z_+}X \Psi - 8 \Bl \g \frac {\z_+}{X^3} \E - \frac{\hX_+}{X^2}
\tB \Br,
 \label{eq:dP}
\end{equation}
where we have combined the parameters $\xi$ and $\k$ into a new
parameter $\g = 6\xi \k^{-2}$. The coefficient in front of $\Psi$
is the logarithmic derivative of the function $\z_+/X^3$, hence if
we define
\begin{equation}
\Psi = \frac {\z_+}{X^3} F,
 \label{eq:PtoF}
\end{equation}
we obtain
\begin{equation}
\frac {dF}{d\z} = -8 \Bl \g\E - \frac{X\hX_+}{\z_+} \tB \Br.
 \label{eq:dF}
\end{equation}
Besides that we have
\begin{equation}
\frac {d\E}{d\z} = \k^2 \Bl \frac F{2X^2} -  \tB \Br, \quad \tB =
-\frac 16 \Bl \k^2 \frac F{8\hX_+} + \frac X{\z_+} \E \Br.
 \label{eq:dEB}
\end{equation}
After inserting for $\tB$ into the expressions for $dF/d\z$ and
$d\E/d\z$ we arrive at the desired system of equations of first
order. We could proceed further and exclude $\E$ to obtain an
equation of second order for $F$, but we will not do that since
the equation will be not needed in what follows.

\subsection{Long-wave limit}

Suppose the shear modulus of the radiation-like solid is a small
fraction of the total energy density of radiation, $|\xi| =
|\mu|/\r_r \ll 1$, and consider a perturbation that is stretched
far beyond the particle horizon at the moment $\eta_*$, $\k =
k\eta_* \ll 1$. We will be mainly interested in the size of the
perturbation at recombination. At that moment, the scale parameter
is $a_{re} = a_{eq} \z_{re} (2 + \z_{re}) \doteq 3 a_{eq}$,
therefore the value of $\z$ is $\z_{re} = \eta_{re}/\eta_* \doteq
1$ and the condition $\k \ll 1$ means that the perturbation is
stretched far beyond the particle horizon {\it at recombination}.
Note that since the value of $c_{S\|}$ does not fall too much
before recombination, such perturbations can be called
``superhorizon'' in the previous sense ``stretched far beyond the
sound horizon''. We will calculate the functions $F$ and $\E$ for
superhorizon perturbations in the leading order in $\xi$ as well
as $\k$, keeping the parameter $\g$ arbitrary.

Suppose the radiation-like solid appeared in the universe at some
moment $\eta_s$ deep in the radiation dominated era, $\eta_s \ll
\eta_{eq}$. For $\eta_s \le \eta \ll \eta_{eq}$ our $\Psi$ must
coincide within a good accuracy with $\Psi$ in the theory with
one-component matter with $w = 1/3$, hence at $\eta \ll \eta_{eq}$
our $\Psi$ must be given by equation (\ref{eq:asP}). By using $\te
= \z/\z_s$ and $z_s = \bar c_{S\|} \k \z_s$, where $\bar c_{S\|}$
is the longitudinal sound speed of radiation combined with
radiation-like solid, $\bar c_{S\|}^2 = 1/3 + \xi$, we find
\begin{equation}
\Psi = \z_s^m [P \z^{-m} + \k^{-2} \tQ (\z^{-2 - m} - \z_s^{2n}
\z^{-2 - M})],
 \label{eq:asPx}
\end{equation}
where $\tQ = {\bar c}_{S\|}^{-2} Q$. The expression for $\Psi$ in
the leading order in $\xi$ is obtained by putting $n = 1/2$, $m =
0$ and $M = 1$ here as well as in the expressions for $p$ and $r$,
and $m = 6\xi$ in the expression for $q$. The resulting formulas
for $p$, $q$ and $r$ yield approximate expressions for $P$ and
$\tQ$ in terms of $\Pz$, which can be interpreted now as the value
of $\Phi$ at the beginning of Friedmann expansion. The expressions
are $P = \Pz$ and $\tQ = 3Q = 27\xi \Pz$, so that
\begin{equation}
\Psi = \BL 1 + \frac 92 \g (\z^{-2} - \z_s \z^{-3}) \BR \Pz.
 \label{eq:asPap}
\end{equation}
Our aim is to extend this expression to the moment of
recombination.

Equations for $F$ and $\E$ in the leading order in $\xi$ and $\k$
are
\begin{equation}
\frac {dF}{d\z} = -8 \Bl \g + \frac{X^2 \hX_+}{6X_+} \Br \E, \quad
\E = const.
 \label{eq:dFdE}
\end{equation}
(The correction to $\E$ is of order $\k^2$, therefore even if $\g
\gg 1$, its contribution to $dF/d\z$ is of order $\xi$ and can be
neglected.) For $\z \ll 1$ the second term in the brackets in the
expression for $dF/d\z$ equals $(2/3) \z^2$, so that
\begin{equation*}
\Psi = \frac 18 \z^{-3} F = - \z^{-3} \int \Bl \g + \frac 23 \z^2
\Br  \E d\z = - \Bl \g \z^{-2} + \frac 29 \Br \E + const \times
\z^{-3},
\end{equation*}
and by comparing this with equation (\ref{eq:asPap}) we rediscover
the formula for $\E$ in a universe with pure radiation, used
already in the matching conditions at the moment of
solidification,
\begin{equation}
\E = - \frac 92 \Pz.
 \label{eq:Eap}
\end{equation}
We can also see that the integration constant in $F$ must be
chosen in such a way that $F = - 8 [\g (\z - \z_s) + O(\z^3)] \E$
for $\z \to 0$. By elementary integration we obtain
\begin{equation}
F = - 8 \BL \g (\z - \z_s) + \frac 1{24} Z \BR \E, \quad Z = \z^3
\Bl\frac 35 \z^2 + 3\z + \frac {13}3 + \frac 1{\z_+}\Br.
 \label{eq:Fap}
\end{equation}
We can now insert for $F$ and $\E$ into the expressions for $\Psi$
(equation (\ref{eq:PtoF})) and $\dP$ (second equation in
(\ref{eq:PPxi})) to arrive at the approximate formulas
\begin{equation}
\Psi = \frac {\z_+}{X^3} \BL 36 \g (\z - \z_s) + \frac 32 Z \BR
\Pz, \quad \dP = - \frac {18\g}{X^2} \Pz.
 \label{eq:PPap}
\end{equation}

The second part of $\Psi$ is the Newtonian potential of
superhorizon perturbations in a universe filled with ideal fluid,
\begin{equation}
\Pid = \frac 32 \frac {\z_+}{X^3} Z \Pz.
 \label{eq:Pid}
\end{equation}
This coincides with the non-decaying part of $\Pid$ given in
equation (7.71) in \cite{mukh}. The function $\Pid$ is constant
both at $\z \ll 1$ (radiation dominated era) and $\z \gg 1$
(matter dominated era), and its value decreases during the
transition between the eras by the factor $\o_\infty = 9/10$. In
approximate calculations, one identifies $\o_\infty$ with the
ratio of the values of $\Pid$ at recombination and at the
beginning of Friedmann expansion \cite{mukh}. However, the actual
ratio is greater. Its value, obtained by inserting $\z = 1$ into
the expression in front of $\Pz$ in (\ref{eq:Pid}), is $\o =
253/270 = 0.94$.

To compute the values of $\Phi$ and $\Psi$ at recombination, we
must insert $\z = 1$ into the expressions for $\Psi$ and $\dP$ in
(\ref{eq:PPap}). We can neglect $\z_s$ since we have assumed
$\eta_s \ll \eta_{eq}$, which implies $\eta_s \ll \eta_*$ and
$\z_s = \eta_s/\eta_* \ll 1$. If we return from $\g$ back to $\xi
\k^{-2}$, we have
\begin{equation}
\Phi_{re} = (4\xi \k^{-2} + \o) \Pz, \quad \Psi_{re} = (16\xi
\k^{-2} + \o) \Pz.
 \label{eq:PPre}
\end{equation}

The potentials $\Phi_{re}$ and $\Psi_{re}$ expressed in terms of
the potential $\Phi_{id,re} = \o \Pz$ are depicted as functions of
$\k$ in fig. 1.
\begin{figure}[ht]
\centerline{\includegraphics[width=0.9\textwidth]{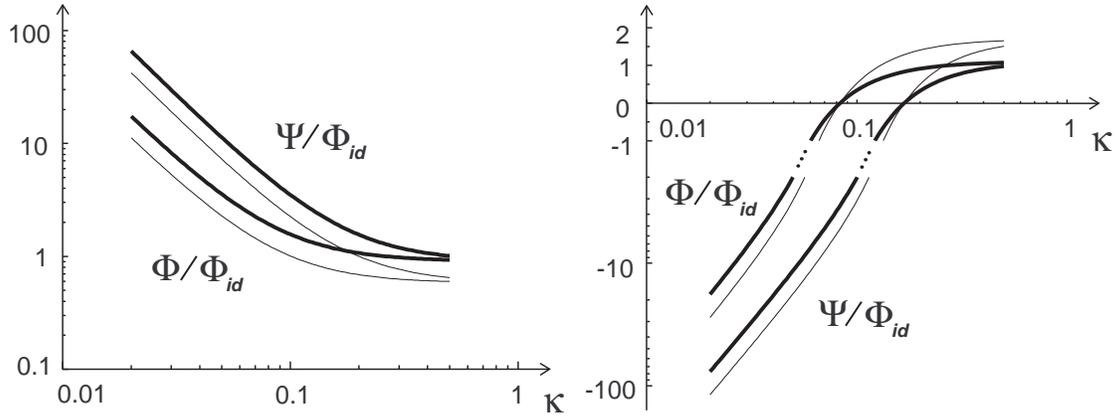}}
\centerline{\parbox {12.2cm}{\caption{\small Gravitational
potentials at recombination as functions of the wave number for
positive (left) and negative (right) shear modulus}}}
  \label{fig:pre}
\end{figure}
The value of the dimensionless shear modulus is $\xi = 1/600$ in
the left panel and $\xi = -1/600$ in the right panel. (The values
are the same as in fig. 1 in \cite{bal}, where the parameter $b =
6\xi$ was used instead of $\xi$.) When constructing the graphs, we
have used a more accurate theory than the one explained in the
text, with the initial Newtonian potential modified to $\hat
\Phi^\0 = 1/(2n) (\z_s/2)^m \Pz$. Heavy curves are computed for
$\z_s = 10^{-5}$, which corresponds to shear stress appearing on
the scale of nucleosynthesis, and light curves are computed for
$\z_s = 10^{-24}$, which corresponds to shear stress appearing on
the GUT scale. The curves cross zero in case $\xi < 0$, therefore
we have combined ordinary and logarithmic scale on the vertical
axis in the right panel.

The expressions for $\Phi_{re}$ and $\Psi_{re}$ were derived under
the condition $\k \ll 1$. In the figure, we have extrapolated the
curves up to $\k = 0.5$, where the error is of order 25\%. At such
$\k$, $\Phi_{re}$ and $\Psi_{re}$ decrease in the exact theory due
to the inhomogeneities being suppressed inside the horizon during
the first quarter-period of acoustic oscillations. In our
approximation the functions are saturated close to $\k = 0.5$,
where the right hand side of equations (\ref{eq:PPre}) is
dominated by the second (constant) term. The first term is small
because of small $|\xi|$. As $\k$ decreases, the first term
becomes dominant and both functions start to rise or fall steeply
depending on the sign of $\xi$.

An alternative description of scalar perturbations to the one used
here is by means of the {\it curvature perturbation} $\R$
\cite{ll}. It is defined as the 3-space curvature in comoving
slicing ${}^{(3)}R_{com}$ with suppressed factor $4(k/a)^2$, and
by using ${}^{(3)}R = - 4(k/a)^2 \psi$ (follows from the
definition) and $\psi_{com} = \H \B$ (is obtained by the time
shift $\d \eta = \B$), we find that it is given by $\R = -\H \B$.
For superhorizon perturbations the first equation (\ref{eq:PP0})
yields $\H \B = -(1/3) \E$, and if we insert here the value $\E =
-(9/2) \Pz$ from equation (\ref{eq:Eap}), we obtain $\R = -(3/2)
\Pz$ irrespective of the value of $|\xi|$ as long as it is small.
Thus, $\R$ is constant in the presence of solid just like in a
universe with pure fluid. However, the large-angle CMB
anisotropies do not inherit the flat spectrum of primordial
perturbations because of that, since they are no longer given
solely by $\R$.

\subsection{Perturbations in the presence of cosmological constant}

For the description of the integrated Sachs-Wolfe effect we need
an extended version of the theory, with the cosmological constant
(dark energy) added to radiation and matter as the third component
of the universe. Thus, we need to consider a universe with the
density
\begin{equation}
\r = \r_{r,eq} (\X^{-4} + \X^{-3} + x_0^{-3}), \quad x_0 = \Bl
\frac {\O_m}{\O_\L} \Br^{1/3} \X_0,
 \label{eq:rg}
\end{equation}
where $\X = a/a_{eq}$, $\O_m$ and $\O_\L$ are the density
parameters of matter and cosmological constant and $\X_0$ is the
present value of $\X$, $\X_0 = a_0/a_{eq} = (\r_m/\r_r)_0 = 24800
\O_m h^2$. By using $\O_m = 0.32$ and $h = 0.67$ ({\it Planck}
values) we find $\X_0 = 3560$, and if we insert this, the cited
value of $\O_m$ and $\O_\L = 1 - \O_m$ into the definition of
$x_0$, we obtain $x_0 = 2770$.

The function $\tH$ can be determined from the first equation
(\ref{eq:Eeq}), after rewriting it as
\begin{equation*}
\tH \equiv \frac 1\X \frac {d\X}{d\z} = \eta_* a_{eq} \Bl \frac 16
\r \X^2 \Br^{1/2}.
\end{equation*}
Clearly, the effect of cosmological constant is significant only
for $\X \gtrsim x_0$, and is completely negligible in the first
period after the moment $\eta_*$ when $\X$ is of order 1. Thus, we
can put $\X = X$ and $\r = \r_{r,eq} (X^{-4} + X^{-3})$ during
this period to obtain
\begin{equation*}
\eta_* a_{eq} \Bl \frac 16 \r_* \Br^{1/2} = \frac 49, \quad \r_* =
\frac 4{81} \r_{r,eq},
\end{equation*}
and
\begin{equation}
\tH = \frac {2\sqrt{\Y}}\X, \quad \Y = 1 + \X + x_0^{-3} \X^4.
 \label{eq:Hg}
\end{equation}

The theory of perturbations we have developed so far works only if
all components of matter are coupled to each other. This is not
true even before recombination, because not all the matter is
coupled to radiation; baryonic matter is, but dark matter is not.
To account for that, one must extend equations for $\B$ and $\E$
by terms containing the 3-space part of the 4-velocity of the dark
matter ${\bf u}_d = a^{-1}\d {\bf x}_d'$, and add to the theory
Euler equation written for dark matter only (equation for $\B$
with zero sound speed) to fix ${\bf u}_d$. Previous equations are
valid only under the simplifying assumption that the dark matter
moves in the same way as the baryon-radiation plasma, ${\bf u}_d =
0$.

We are interested in the evolution of perturbations in the period
after recombination, when baryonic matter, too, is decoupled from
radiation. Radiation then acquires nonzero velocity ${\bf u}_r =
a^{-1}\d {\bf x}_r'$, but its evolution looks different than that
of dark matter. It undergoes {\it free-streaming}, described by
equation (\ref{eq:dDTP}) from the next section. As we will see, an
approximate theory of free streaming based on this equation gives
not quite as large an effect as we would need. Thus, if we do not
aim at a complete description of perturbations and just want to
estimate the effect of a radiation-like solid on them, we can
suppose that the matter filling the universe is fully coupled
after recombination and account for the effect of decoupling, as
well as for any other effect that is possibly important and was
not taken into consideration, by an {\it ad hoc} correction of the
part of the theory determining the CMB anisotropies in the
presence of pure fluid.

In a universe with fully coupled matter, the auxiliary sound speed
squared is
\begin{equation}
c_{S0}^2 = \frac {K_r}{\r_+} = \frac {(1/3) \r_{r+}}{\r_{r+} +
\r_m} = \frac {1/3}{\hcX_+},
 \label{eq:cS2}
\end{equation}
where $\hcX_+$ is defined in the same way in terms of $\X$ as
$X_+$ is defined in terms of $X$, $\hcX_+ = 1 + (3/4)\X$.
(Cosmological constant does not contribute to $c_{S0}^2$, since
$\r_{\L+} = \r_\L + p_\L = 0$.) Furthermore, for the contribution
of the solid to $c_{S\|}^2$ and the function $\a$ we have
\begin{equation*}
\D c_{S\|}^2 = \frac 43 \frac {\xi \r_r}{\r_+} = 3\xi c_{S0}^2,
\quad \a = \frac 32 \frac 1{3c_{S0}^2} \frac {\r_{r+}}\r = \frac
2{3c_{S0}^2 \Y}.
\end{equation*}

Our starting point will be equations (\ref{eq:dBdE}) and
(\ref{eq:PP0}) with the prime replaced by $d/d\z$, $k$ replaced by
$\k$ and tilde attached to $\B$ and $\H$ as well as to $\xi$ in
the expression for $\dP$. First we find the generalized version of
equation (\ref{eq:dP}),
\begin{equation}
\frac {d\Psi}{d\z} = \BL \frac 1{\a \tH^2} \frac {d(\a
\tH^2)}{d\z} + (3c_{S0}^2 - \a) \tH \BR \Psi - \a \tH^2 \Bl \frac
32 \g c_{S0}^2 \tH \E - \tB \Br.
 \label{eq:dPg}
\end{equation}
By using the second equation in (\ref{eq:Eeq}) along with $p' =
-3\H K$ and the first equation (\ref{eq:Eeq}) along with $\H' = -3
(\r + 3p)a^2$, we obtain
\begin{equation*}
\frac 1\a \frac {d\a}{d\z} = \frac 1{\r_+} \frac {d\r_+}{d\z} -
\frac 1\r \frac {d\r}{d\z} = 3\tH \Bl \frac p\r - \frac K{\r_+}\Br
= [-3(1 + c_{S0}^2) + 2\a] \tH, \quad \frac 1{\tH^2} \frac
{d\tH}{d\z} = 1 - \a,
\end{equation*}
so that the expression in front of $\Psi$ in equation
(\ref{eq:dPg}) can be written as
\begin{equation*}
\frac 1{\a \tH^2} \frac {d(\a \tH^2)}{d\z} + (3c_{S0}^2 - \a) \tH
= - (1 + \a) \tH = - 2\tH + \frac 1\tH \frac {d\tH}{d\z} = \frac
1{\X^{-2} \tH} \frac {d(\X^{-2} \tH)}{d\z}.
\end{equation*}
As a result, if we define
\begin{equation}
\Psi = \frac \tH{2\X^2} F,
 \label{eq:PtoFg}
\end{equation}
and use $\a \tH^2 = 8/(3c_{S0}^2 \X^2)$, we arrive at
\begin{equation}
\frac {dF}{d\z} = - 8 \Bl \g \E - \frac 2{3 c_{S0}^2 \tH} \tB \Br.
 \label{eq:dFg}
\end{equation}
This is the generalized version of equation (\ref{eq:dF}).
Analogically, for the functions $\E$ and $\tB$ we have the
generalized version of equations (\ref{eq:dEB}),
\begin{equation}
\frac {d\E}{d\z} = \k^2 \Bl \frac F{2\X^2} -  \tB \Br, \quad \tB =
-\frac 16 \Bl \k^2 \frac {3 c_{S0}^2}8 F + \frac 2{\tH} \E \Br.
 \label{eq:dEBg}
\end{equation}
We also find, after writing the modified parameter $\xi$ as $\tx =
(3/2) \xi c_{S0}^2 \a$, that the generalized version of the second
equation in (\ref{eq:PPxi}) is $\dP = (24\xi/\X^2) \k^{-2} \E =
(4\g/\X^2) \E$. We can see that the new equations reduce to the
old ones with the replacements $X \to \X$, $\hX_+ \to \hcX_+$ and
$\xi_+ \to \sqrt{\Y}$. As for the function $\X(\z)$ appearing in
them, we do not have an analytical expression for it anymore. It
must be calculated numerically from equation (\ref{eq:Hg}),
rewritten as $d\X/d\z = 2\sqrt{\Y}$.

In the long-wave limit we have equations (\ref{eq:dFdE}) with the
replacements listed above for $F$ and $\E$ and expression
(\ref{eq:Eap}) for $\E$. The resulting generalization of equations
(\ref{eq:PPap}) is
\begin{equation}
\Psi = 36 \frac {\sqrt{\Y}}{\X^3} \BL \g (\z - \z_s) + \int
\frac{\X^2 \hcX_+}{6\Y} d\z \BR \Pz, \quad \dP = - \frac
{18\g}{\X^2} \Pz,
 \label{eq:PPapg}
\end{equation}
The second part of $\Psi$ is the potential $\Pid$ in the presence
of cosmological constant. It can be transformed into the form
given in equation (7.69) in \cite{mukh} by using the identity
\begin{equation*}
\int \frac{\X^2 \hcX_+}\Y d\z = \frac 14 \Bl \frac
{\X^3}{\sqrt{\Y}} - 2 \int \X^2 d\z \Br,
\end{equation*}
which can be obtained by integration by parts. Again, we are
restricting ourselves to the nondecaying part of $\Pid$. This can
be done by choosing the integration constant in such a way that
the integral is of order $\z^3$ for $\z \to 0$; or equivalently,
by regarding the integral as definite, starting from $\z = 0$.


\section{Large-angle CMB anisotropies}
 \label{sec:eff}

\subsection{Temperature fluctuations}

To compute CMB anisotropies we must know the relative deviation of
the temperature from its mean value $\D T = \d T/T$ at the time of
recombination. Since $\r_r \propto T^4$, it holds
\begin{equation}
\D T =\frac 14 \d_\g,
 \label{eq:dT}
\end{equation}
where $\d_\g$ is the density contrast of radiation, $\d_\g =
\overline{\d \r}_r/\r_r$. Density perturbations of individual
components of matter, if independent from each other, are given by
a formula analogical to that for the total density perturbation,
$\overline{\d \r}_i = \r_{i+} (3\Psi + \E)$. For the density
contrast $\d_\g$ this yields
\begin{equation}
\d_\g = 4 \Bl \Psi + \frac 13 \E \Br.
 \label{eq:dg}
\end{equation}

Instead of $\D T$ we need the {\it effective} fluctuation of
temperature $\D T_{eff}$, measured by a local observer that is at
rest with respect to the unperturbed matter and observes the
radiation arriving in the same direction in which it then
propagates to the observer on Earth. Let $\bf v$ be the local
velocity of matter, ${\bf v} =\d {\bf x}'$, and $\bf l$ be the
unit vector in the direction of propagation of radiation.
Introduce also the unit vector $\bf n$ pointing from Earth towards
the place on the sky from which the radiation is coming. The
function $\D T_{eff}$ differs from $\D T$ by the Doppler term $\D
T_D = {\bf v} \cdot {\bf l} = - {\bf v} \cdot {\bf n}$. Suppose
the velocity reduces to its scalar part, ${\bf v} = {\bf
v}^{(S)}$, and denote the unit vector in the direction of $\bf k$
by $\bf m$. By performing the transformation from the proper-time
comoving gauge to the Newtonian gauge we find ${\bf v} = - i {\bf
k} E' = i {\bf m} \k^{-1} d\E/d \z$ and
\begin{equation}
\D T_D = - i {\bf m} \cdot {\bf n} \k^{-1} \frac {d\E}{d \z}.
 \label{eq:dTD}
\end{equation}
Note that the expression for $\bf v$ can be obtained also from the
energy conservation law,
\begin{equation*}
(\d_\g - 4\Psi)' + \frac 43 \nabla \cdot {\bf v} = 0,
\end{equation*}
if one inserts into it from equation (\ref{eq:dg}). The law is the
same as in the theory with an ideal fluid, see equation (7.110) in
\cite{mukh}, except that $\Phi$ is replaced by $\Psi$.

CMB anisotropies are calculated from the temperature fluctuations
$\D T_0$ seen today on Earth. (The index `0' denotes the present
moment.) To compute $\D T_0$ we make use of the fact that the
temperature fluctuations of the radiation propagating freely from
the surface of last scattering satisfy
\begin{equation}
\frac {d(\D T + \Phi)}{d\eta} = \frac {\partial(\Phi +
\Psi)}{\partial \eta}.
 \label{eq:dDTP}
\end{equation}
Again, this is the same equation as in the theory with an ideal
fluid, except that one function $\Phi$ on the right hand side is
replaced by $\Psi$, see equation (9.20) in \cite{mukh}.

The total temperature fluctuation $\D T_0$ is a sum of two terms,
one obtained from equation (\ref{eq:dDTP}) without the right hand
side and another one contributed by the right hand side. The
mechanisms responsible for these two terms are called {\it
ordinary Sachs-Wolfe effect} and {\it integrated Sachs-Wolfe
effect}. Let us first compute the contribution to $\D T_0$ from
the ordinary SW effect. If we put the value of $\Phi$ measured on
Earth at present equal to zero, we find $\D T_0^{SW} = (\D T_{eff}
+ \Phi)_{re}$; and by inserting here from equations (\ref{eq:dT}),
(\ref{eq:dg}) and (\ref{eq:dTD}) we obtain
\begin{equation}
\D T_0^{SW} = \Bl\Phi + \Psi + \frac 13 \E - i {\bf m} \cdot {\bf
n} \k^{-1} \frac {d\E}{d \z}\Br_{re}.
 \label{eq:dT0}
\end{equation}
For superhorizon perturbations the last term is negligible since,
as seen from the first equation in (\ref{eq:dEB}), it is of order
$\k \Pz$, while the first three terms are of order $\Pz$. After
skipping the former term and inserting for the latter terms from
equations (\ref{eq:PPre}) and (\ref{eq:Eap}), we obtain
\begin{equation}
\D T_0^{SW} = (20 \xi \k^{-2} + b) \Pz,
 \label{eq:dT0app}
\end{equation}
where $b = 2\o - 3/2 = 101/270 = 0.37$. For an ideal fluid the
first term vanishes and the formula reduces to $\D T_0^{SW} =
b\Pz$. In approximate calculations, one replaces $b$ by $b_\infty$
obtained in the limit $\z_{re} \gg 1$, $b_\infty = 2\o_\infty -
3/2 = 3/10$, see par. 9.5 in \cite{mukh}.

Let us now determine the contribution to $\D T_0$ from the
integrated SW effect. For any function $f$ describing the
perturbation one must distinguish between the amplitude $\tilde f
(\eta)$ and the complete wave $f ({\bf x}, \eta) = \tilde f
e^{i{\bf k} \cdot {\bf x}}$. If we denote the sum $\D T + \Phi$ as
$\tau$, equation (\ref{eq:dDTP}) can be written as an ordinary
differential equation for the function $\tilde \tau$,
\begin{equation*}
\tilde \tau' + ik_\| \tilde \tau = \tilde \Phi' + \tilde \Psi',
\end{equation*}
where $k_\|$ is the projection of the vector $\bf k$ into the
direction of propagation of radiation, $\kp = {\bf k} \cdot {\bf
l} = - {\bf k} \cdot {\bf n}$. The solution is
\begin{equation*}
\tilde \tau e^{i\kp \eta} = \int (\tilde \Phi' + \tilde \Psi')
e^{i\kp \eta} d\eta,
\end{equation*}
so that
\begin{equation}
\tilde \tau_0 = \tilde \tau_{re} e^{-i\kp(\eta_0 - \eta_{re})} +
\int \limits_{\eta_{re}}^{\eta_0} (\tilde \Phi' + \tilde \Psi')
e^{-i\kp(\eta_0 - \eta)} d\eta.
 \label{eq:tM0}
\end{equation}

The value of $\tau$ at present is $\tau_0 = \tilde \tau_0 e^{i
{\bf k} \cdot {\bf x}_0} = \tilde \tau_0$ (we suppose that the
observer is located at the origin) and the value of $\tau$ at
recombination is $\tau_{re} = \tilde \tau_{re} e^{i {\bf k} \cdot
{\bf x}_{re}} = \tilde \tau_{re} e^{-i\kp(\eta_0 - \eta_{re})}$.
On the other hand, from the definition of $\tau$ we have $\tau_0 =
\D T_0$ and $\tau_{re} = (\D T + \Phi)_{re}$; or, if we include
the contribution of Doppler effect to the fluctuation of
temperature, $\tau_{re} = (\D T_{eff} + \Phi)_{re}$. Thus,
equation (\ref{eq:tM0}) can be written as $\D T_0 = \D T_0^{SW} +
\D T_0^{iSW}$, with $\D T_0^{SW}$ given in equation (\ref{eq:dT0})
and
\begin{equation}
\D T_0^{iSW} = \int \limits_{\eta_{re}}^{\eta_0} (u' + v')
e^{i{\bf k} \cdot {\bf n} (\eta_0 - \eta)} d\eta \tilde \Phi^\0,
 \label{eq:DTiSW}
\end{equation}
where we have denoted the functions $\Phi$ and $\Psi$ with
suppressed factor $\Pz$ as $u$ and $v$. With the notation
introduced here we can also repair equation (\ref{eq:dT0app}): the
initial Newtonian potential appearing there must be shifted to the
time $\eta_{re}$, $\Pz \to e^{i{\bf k} \cdot {\bf n}(\eta_0 -
\eta_{re})} \tilde \Phi^\0$.

We have described the integrated SW effect in the framework of
linearized theory. The full theory accounts also for the fact that
the light in the late period propagates through {\it large}
fluctuations of density (superclusters and supervoids), for which
such description is not applicable. The resulting effect, called
{\it nonlinear integrated SW effect}, will not be considered here.

To complete the discussion, let us look how the decoupling of
radiation from baryonic matter influences the evolution of the
potentials $\Phi$ and $\Psi$. Consider first the density contrast
of radiation $\d_\g$. By solving equation (\ref{eq:dDTP}) without
gravitational potentials and averaging the resulting function $\D
T ({\bf x}, {\bf l}, \eta)$ over the directions of $\bf l$, we
obtain
\begin{equation}
\d_\g = - \b j_0 (z) \Pz,
 \label{eq:drr}
\end{equation}
where $\b = 4(3/2 - \o) = 304/135 = 2.25$ and $z = k(\eta -
\eta_{re})$; see equation (9.23) in \cite{mukh}. (We have used the
expression for the density contrast of radiation at recombination
$\d_{\g,re} = - \b \Pz$.) This is to be compared with the density
contrast of radiation in a universe with fully coupled matter,
\begin{equation}
\d_\g^{(c)} = 4\Bl v - \frac 32 \Br \Pz.
 \label{eq:drrc}
\end{equation}
Both functions start from $\d_{\g,re} = - \b \Pz$, but then the
second function mildly increases in absolute value until it
reaches its present value 1.40 $\d_{\g,re}$, while the first
function has three different regimes depending on the parameter
$z_0 \doteq k\eta_0$: for $z_0 \ll 1$ it is practically constant,
for $z_0 \sim 1$ it monotonically decreases in absolute value and
for $z_0 \gg 1$ it oscillates with decreasing amplitude.

In a universe with several decoupled components every component
has its own sound speed. However, even then one can introduce an
effective common sound speed $c_{S,eff} = (\overline{\d
p}/\overline{\d \r})^{1/2}$, whose square appears in the
differential equation of second order for $\Psi$ (or $\Phi$ if the
universe is filled with pure fluid) and in such a way determines
the time dependence of both $\Phi$ and $\Psi$ \cite{mukh}. If the
perturbations are adiabatic and the matter is fully coupled, the
effective sound speed reduces to $[(dp/d\r)_S]^{1/2}$, which is
the actual sound speed in pure fluid and auxiliary sound speed in
a matter containing solid component. In general, the effective
sound speed is given by
\begin{equation*}
c_{S,eff}^2 = \frac {(1/3) \overline{\d \r}_r}{\overline{\d \r}_r
+ \overline{\d \r}_m} = \frac {(1/3) \d_\g}{\d_\g + \d_m \X},
\end{equation*}
and if we express $\d_m$ in terms of $\d_\g^{(c)}$ by using the
formula $\d_i = (\r_{i+}/\r_i) (3\Psi + \E)$, we obtain
\begin{equation}
c_{S,eff}^2 = \frac {(1/3) \d_\g}{\d_\g + (3/4) \d_\g^{(c)} \X}.
 \label{eq:cSe2}
\end{equation}
Clearly, for fully coupled matter this coincides with the
expression (\ref{eq:cS2}) for $c_{S0}^2$. If the radiation is
decoupled, the values of $c_{S,eff}^2$ become smaller, and for
large enough wave numbers they can even cross zero and start to
oscillate. This causes the function $\Psi$ to decrease a bit
slower in the first period after recombination, but the effect is
at most 10 to 20\% for physically interesting values of
parameters.

In addition to reducing the value of $c_{S,eff}^2$, free-streaming
also produces anisotropic stress, hence $\D \Phi$ becomes nonzero
after recombination even in the absence of solid. However, this
effect, if estimated by using the same approximate function $\D T
({\bf x}, {\bf l}, \eta)$ as in the computation of $\d_\g$, turns
out to be negligible.

\subsection{Power spectrum}

The power spectrum of CMB anisotropies are the coefficients $C_l$
in the expansion of two-point correlation function of $\D T_0$,
\begin{equation}
C(\th) \equiv \Big\langle \D T_0({\bf n}) \D T_0({\bf
n}')\Big\rangle = \frac 1{4\pi} \sum (2l + 1) C_l P_l(\cos \th),
 \label{eq:dfC}
\end{equation}
where the angle brackets denote averaging over different regions
in the universe, $P_l$ are Legendre polynomials and $\th$ is the
angle between $\bf n$ and ${\bf n}'$.
(We will use this definition, although when presenting
observational data one usually regards $C_l$ as {\it dimensional}
quantities, defined in terms of $\d T_0$ rather than $\D T_0$.) To
compute $C_l$ for small $l$, we need to know $\D T_0$ for small
$\k$. For the time being, we will restrict ourselves to $\D T_0$
coming from the ordinary SW effect and postpone the discussion of
the integrated SW effect to the next subsection. Thus, we will
identify $\D T_0$ with $\D T_0^{SW}$ of equation
(\ref{eq:dT0app}). What we need to compute $C_l$ is, however, not
what appears in that equation. The quantity $\D T_0$ on its left
hand side is actually the {\it Fourier coefficient} of the
deviation of temperature from the mean value, and the quantity
$\Pz$ on its right hand side is the {\it Fourier coefficient} of
the initial Newtonian potential. In other words, the equation must
be read as
\begin{equation}
\D T_{0{\bf k}} = (20 \xi \k^{-2} + b) e^{i {\bf k} \cdot {\bf n}
\eta_0} \Pz_{\bf k}.
 \label{eq:dT0k}
\end{equation}
The exponential in front of $\Pz_{\bf k}$, mentioned already in
the discussion after equation (\ref{eq:DTiSW}), takes into account
the fact that the right hand side of equation (\ref{eq:dT0app})
refers to the moment of recombination. We should actually write
there $e^{i {\bf k} \cdot {\bf n} (\eta_0 - \eta_{re})}$, but the
time $\eta_{re}$ is about 50 times less than the time $\eta_0$,
therefore we have neglected it.

We want to calculate the coefficients $C_l$ in order to establish
what values of $\xi$ are allowed by observations. Since our
transfer function is not constant but rises with decreasing wave
number, we expect $C_l$ to rise with decreasing multipole moment;
and $|\xi|$ must not be too large in order that this behavior is
in agreement with observations within cosmic variance. Besides
$C_l$, one observes also the {\it bi\-spectrum} $B_{l_1 l_2 l_3}$,
which encodes information about the three-point correlation
function of $\D T_0$. Nonzero coefficients $B_{l_1 l_2 l_3}$ mean
non-Gaussian probability distribution of $\D T_0$, therefore by
measuring the bi\-spectrum one can obtain an observational upper
limit on the parameter of non-Gaussianity $f_{NL}$. In such a way
one constrains the parameter space of alternative inflationary
scenarios producing nonzero $f_{NL}$, like one-field inflation
with non-standard kinetic term or multifield inflation with highly
nonlinear energy-momentum tensor of one field \cite{sasa}. The
behavior of our transfer function suggests that solidification of
a part of radiation in the early universe leads to the enhancement
of the coefficients $B_{l_1 l_2 l_3}$ for small multipole moments.
Thus, the parameter $\xi$ can be constrained also by observational
data on the bispectrum.

From inflation one obtains a spectrum of $\Pz$ which is, within a
good accuracy, {\it flat}; that is, $\langle  \Pz_{\bf k}
\Phi^{(0)*}_{{\bf k}'} \rangle = B k^{-3} \d({\bf k} - {\bf k}')$
with constant $B$. By inserting $\D T_0 ({\bf n}) = \displaystyle
\int \D T_{0{\bf k}} \frac {d^3k}{(2\pi)^{3/2}}$ into
(\ref{eq:dfC}) and using the expression for $\langle \Pz_{\bf k}
\Pz_{{\bf k}'} \rangle$, we find
\begin{equation*}
C_l = \frac 2\pi B \int \limits_0^\infty (20 \xi \k^{-2} + b)^2
j_l^2 (k\eta_0) \frac {dk}k,
\end{equation*}
where $j_l$ is the spherical Bessel function. Next we pass from
$k$ to $s = k\eta_0$ to obtain
\begin{equation}
C_l = \frac 2\pi Bb^2 \int \limits_0^\infty (\xs s^{-2} + 1)^2
j_l^2 (s) \frac {ds}s,
 \label{eq:Cint}
\end{equation}
where $\xs = (20/b) (\eta_0/\eta_*)^2\xi$. Finally we compute the
integral with the help of the formula
\begin{equation*}
\int \limits_0^\infty s^{-n} j_l^2 \frac {ds}s = \frac \pi{8 \cdot
2^n} \frac{\G(2 + n)}{\G^2(\tfrac{3 + n}2)} \frac{\G(l - \tfrac
n2)}{\G(l + 2 + \tfrac n2)},
\end{equation*}
to find
\begin{equation}
C_l = \BL \frac 8{15} \frac {\xs^2 }{(l + 3)(l + 2)(l - 1)(l - 2)}
+ \frac 43 \frac {\xs}{(l + 2)(l - 1)} + 1\BR C_{l,id},
 \label{eq:C}
\end{equation}
where
\begin{equation}
C_{l,id} = \frac 1{l(l + 1)}\frac {Bb^2}{\pi}.
 \label{eq:Cid}
\end{equation}
After $b$ is replaced by $b_\infty$, the expression for $C_{l,id}$
coincides with that in equation (9.44) in \cite{mukh}.

To complete the theory, we need the numerical value of the
coefficient of proportionality between $\xs$ and $\xi$. By
integrating the equation $d\X/d\z = 2\sqrt{\Y}$ from $\z = 1$, $\X
= 3$ to the value of $\z$ at which $\X = \X_0$, we obtain $\z_0 =
53.4$, and after inserting this into the definition of $\xs$ for
$\eta_0/\eta_*$ we find
\begin{equation}
\xs = 1.52 \times 10^5 \xi.
 \label{eq:xs}
\end{equation}

The coefficients $C_0$, $C_1$ and $C_2$ are infinite, first of
them even in the case when the cosmic medium is fluid. (This is
true for $C_0$ unless $\xs = (15/2) (1 \pm \sqrt{3/5})$ and for
$C_1$ unless $\xs = 10$. However, as seen from the following
discussion, there is no need to explore these singular cases in
detail.) The coefficients become finite if we take into account
that the spectrum of perturbations is cut off at some wave number
$k_{min}$ given by the duration of inflation; and the coefficient
$C_2$, as well as the coefficient $C_0$ in case $\xi = 0$, become
finite also if we introduce a small negative tilt of the
primordial spectrum, replacing $B$ by $B s^\e$ with $\e > 0$.
However, of the three coefficients we need to consider $C_2$
(quadrupole) only. The other two coefficients (monopole and
dipole) do not enter the theory, because an individual observer
has no clue how much the mean temperature $T_{obs}$ he has
measured differs from the true mean temperature $T$, neither can
he tell how much his velocity with respect to CMB $V_{obs}$, which
he determines by averaging the product $\D T_{obs} \cos \th$,
differs from his true velocity $V$. He identifies $T$ with
$T_{obs}$ and $V$ with $V_{obs}$; and if we write the coefficients
$C_l$ as
\begin{equation*}
C_l = \Big\langle \D T_0({\bf n}) \int \D T_0({\bf n}') P_l
d\Omega'\Big\rangle,
\end{equation*}
we can see that such identification means that the coefficients
$C_0$ and $C_1$ are put equal to zero.

The constant $c$ appearing in the formula for the renormalized
coefficient $C_2$,
\begin{equation}
C_{2r} = \Bl \u \xs^2 + \frac 13 \xs + 1\Br C_{2id},
 \label{eq:Cren}
\end{equation}
can be written as
\begin{equation}
\u = \frac 4{75} [\e^{-1} (1 - \sm^\e) + d] \doteq \frac 4{75}
\times \bigg\{ \mbox{\hskip -2mm}
  \left. \begin{array} {l}
  \e^{-1} \equiv \u_I \mbox{ if }  \e \log (1/\sm) \gg 1\\
  \log (1/\sm) + d \equiv \u_\I \mbox{ if } \e \log (1/\sm) \ll 1\\
  \end{array}\mbox{\hskip -1.5mm}, \right.
 \label{eq:u}
\end{equation}
where the constant $d$ is given by
\begin{equation*}
d = \log 2 - \frac 12 \psi(6) + \frac 12 \psi(1) + \psi(7/2) =
\frac {77}{40} - \log 2 - \g = 0.655.
\end{equation*}
Here $\psi$ is digamma function and $\g$ is Euler-Mascheroni
constant. The parameter $\e$ is the deviation of the scalar
spectral index $n_S$ from 1, $\e = 1 - n_S$, so that $\e = 0.04$
and $\u_I = 4/3$ for the observational mean value of $n_S$, which
is 0.96 (again a {\it Planck} value). The value of the parameter
$\sm$ depends on the inflationary scenario. It holds
\begin{equation*}
\sm = k_{min} \eta_0 = \frac {r_{h0}}{\lr_{max0}} =
\frac{r_{obs}^{(0)}}{{\lr}_{max}^{(0)}} = \frac {N_{min}
r_{h,inf}}{N H^{-1}} \approx \frac {N_{min}}N,
\end{equation*}
where the index `(0)' denotes the beginning of Friedmann
expansion, $r_{obs}$ is the radius of the part of the universe
which is observable today, the index `inf' denotes the beginning
of inflation, $N$ is the number of $e$-foldings during inflation,
$N_{min}$ is minimum $N$ and $H$ is Hubble constant during
inflation. For inflation on GUT and Planck scale (new and chaotic)
$N$ is typically about 2000 and $10^7 \div 10^{11}$ respectively,
while $N_{min}$ has a value between 60 and 70. This yields $\u_\I
= 0.22$ and $0.67 \div 1.16$, hence $\u_\I \ll \u_I$ for inflation
on the GUT scale and $\u_\I \lesssim \u_I$ for inflation on the
Planck scale. According to (\ref{eq:u}), the value of $\u$ in both
asymptotic regimes is given by the less of the two numbers $\u_I$
and $\u_\I$. Thus, in approximate calculations we can use $\u =
\u_\I$; in other words, we can ignore the tilt of the primordial
spectrum and take into account only its cutoff.

The observed values of $C_l$ must coincide with the theoretical
values within cosmic variance,
\begin{equation}
C_{l,obs} \in \langle 1 - \d_l, 1 + \d_l\rangle  C_l, \quad \d_l =
\sqrt{\frac 2{2l + 1}}.
 \label{eq:var}
\end{equation}
Denote the relative deviation of our $C_{l}$ from $C_{l,id}$ by
$\D_l$. After identifying $C_{l,obs}$ with $C_{l,id}$ we find that
$\xs$ must assume values between $- 4.15$ and 2.30 (a consequence
of $1/(1 + \D_{2r}) \ge 1 - \d_2$, if one inserts for $c$ the
value $c_\I$ computed for inflation on GUT scale). Thus, both
positive and negative values of $\xi$ are admissible and $|\xi|$
cannot exceed values of order $10^{-5}$.

The identification of $C_{l,obs}$ and $C_{l,id}$ is more or less
acceptable for all multipoles except for the quadrupole, whose
observational value lies approximately at the lower limit of the
interval allowed by cosmic variance, $C_{2,obs} \doteq ( 1 - \d_2)
C_{2,id}$. The theory with a solid can provide for that only for
$\xs$ between $- 1.85$ and 0 (a consequence of $\D_{2r}\le 0$),
which is an unwanted result since negative values of $\xi$ are
most probably unphysical and were included into the theory only
for completeness. To return positive values of $\xi$ into the
game, we can use an idea put forward in the cosmology with pure
fluid, that the lack of power in the quadrupole comes from a
cutoff of the primordial spectrum on the scale $\sm \sim 1$
\cite{{bri},{con},{cli}}. Such strong cutoff can be caused either
by short duration of inflation or by jump-like variation of
inflationary potential close to the end of inflation. In the
presence of solid, we can use strong cutoff to reconcile positive
$\xi$ with observations, while explaining small $C_{2,obs}$ as
before by cosmic variance. Suppose the integral in the expression
(\ref{eq:Cint}) is cut at $\sm = 2$ and write the power spectrum
as $C_l = (p_l \xs^2 + 2q_l \xs + r_l) C_{l,id}$, where $C_{l,id}$
is the power spectrum in the theory with pure fluid in which no
cutoff occurs. After imposing the condition that the ratio
$C_l/C_{l,id}$ equals $1 - \d_2$ for $l = 2$ and 1 for $l > 2$
within the cosmic variance, we find that the maximum $\xs$ is 0.81
(a consequence of $p_2 \xs^2 + 2q_2 \xs + r_2 = 1$) and the
minimum $\xs$ is $-3.21$ (a consequence of $p_3 \xs^2 + 2q_3 \xs +
r_3 = 1/(1 + \d_3)$). Again, we conclude that $\xi$ can be
positive as well as negative and must satisfy $|\xi| \lesssim
10^{-5}$.

\subsection{Switching on the integrated SW effect}

The formula for $\D T_{0{\bf k}}$ that takes into account the
contribution of the integrated SW effect is
\begin{equation}
\D T_{0{\bf k}} = \BL f_k e^{i {\bf k} \cdot {\bf n} \eta_0} +
\int \limits_{\eta_{re}}^{\eta_0} g_k' e^{i{\bf k} \cdot {\bf n}
(\eta_0 - \eta)} d\eta \BR \Pz_{\bf k},
 \label{eq:dTtot}
\end{equation}
where $f_k = 20 \xi \k^{-2} + b$ and $g_k = u_k + v_k = (\Phi_{\bf
k} + \Psi_{\bf k})/\Pz_{\bf k}$. Note that we cannot put
$\eta_{re} = 0$ in the lower limit of the integral, like we did in
the argument of the exponential $e^{i {\bf k} \cdot {\bf n}
(\eta_0 - \eta_{re})}$, because the contribution of the solid to
the integral would diverge. For superhorizon perturbations the
function $g_k$ can be extracted from expressions (\ref{eq:PPapg})
for $\Psi$ and $\dP$. The result is
\begin{equation}
g_k = b (\xi_* s^{-2}g_s + g_{id}),
 \label{eq:u0v0}
\end{equation}
where
\begin{equation}
g_s = \frac{108}5 \Bl \frac {\sqrt{\Y}}{\X^3} \z - \frac 1{4\X^2}
\Br, \quad g_{id} = 12 b^{-1} \frac{\sqrt{\Y}}{\X^3} \int
\frac{\X^2 \hcX_+}\Y d\z.
 \label{eq:gsgi}
\end{equation}
We have skipped the shift by $\z_s$ in the expression $\z - \z_s$
appearing in $g_s$, since the functions $g_s$ and $g_{id}$ are
needed only in the interval $\z \ge 1 \gg \z_s$.

The expression for $g_k$ is valid if the perturbation is stretched
far beyond the sound horizon,
\begin{equation}
c_{S0} k \eta = \frac {\eta_*}{\eta_0} s c_{S0} \z \ll 1,
 \label{eq:cond}
\end{equation}
where $c_{S0}$ is given in equation (\ref{eq:cS2}). The period
between the times $\eta_{re}$ and $\eta_0$ consists of two
distinctive eras, the early era dominated by matter and the late
era dominated by cosmological constant. They are separated by the
time $\eta_{eq}'$ at which the densities $\r_m$ and $\r_\L$
coincide, $\eta_{eq}' = 49.6$ for the {\it Planck} values of
cosmological parameters. Note, however, that the effect of
cosmological constant shows up considerably earlier; for example,
the function $g_{id}$ switches from the early-time regime to the
late-time regime as soon as at $\eta \sim 15$. The function $\Z =
c_{S0} \z$ rises during the early period from the value $\Z_{re} =
\sqrt{4/39} = 0.32$ to a value close to $\Z_\infty = 2/3$ (the
limiting value in a universe without cosmological constant), and
slightly recedes during the late period. Thus, equation
(\ref{eq:u0v0}) can be safely used for the values of $s$
substantially smaller than the horizon value $s_S =
(\eta_0/\eta_*) \Z_\infty^{-1} = 80.0$.

We have formulated the condition for long-wave perturbations
assuming, like in the computation of $g_k$, that the matter
filling the universe after recombination is fully coupled. In the
theory with decoupled radiation, the effective sound speed is
smaller and the sound horizon is bigger, hence the value of $s_S$
is even greater.

Rewrite the expression for $\D T_{0{\bf k}}$ into a more
convenient form,
\begin{equation*}
\D T_{0{\bf k}} = \Bl f_k e^{is {\bf m} \cdot {\bf n}} +\int
\limits_{\s_{re}}^1 \frac {dg_k}{d\s} e^{i \bs{\bf m} \cdot {\bf
n}} d\s \Br \Pz_{\bf k},
\end{equation*}
where $\s = \eta/\eta_0$ and $\bs = (1 - \s)s$. After inserting
this into the mean value $\langle \D T_{0{\bf k}} ({\bf n}) \D
T_{0{\bf k}} ({\bf n}') \rangle$ and using the identity
\begin{equation*}
\Bla e^{i{\bf m} \cdot (s{\bf n} - s'{\bf n}')} \Bra_{\bf m} =
\sum (2l + 1) j_l (s) j_l (s') P_l(\cos \th),
\end{equation*}
where $\langle \ \rangle_{\bf m} = \displaystyle \int (\ldots)
\frac {d\O_{\bf m}}{4\pi}$, we obtain
\begin{equation}
C_l = \frac 2\pi B \int \limits_0^\infty [f_k j_l (s) + \cG_l
(s)]^2 \frac {ds}s, \quad \cG_l = \int \limits_{\s_{re}}^1 \frac
{dg_k}{d\s} j_l (\bs) d\s.
 \label{eq:CiSW}
\end{equation}
With the function $f_k$ written as $f_k = b(\xi_* s^{-2} + 1)$ and
the function $g_k$ given in equations (\ref{eq:u0v0}) and
(\ref{eq:gsgi}), this yields
\begin{equation}
C_l = \frac 2\pi B b^2 \int \limits_0^\infty [\xi_* s^{-2} (j_l +
\cG_{l,s}) + j_l + \cG_{l,id}]^2 \frac {ds}s,
 \label{eq:CiSWs}
\end{equation}
where
\begin{equation}
\cG_{l,s} = \int \limits_{\s_{re}}^1 \frac {dg_s}{d\s} j_l (\bs)
d\s, \quad \cG_{l,id} = \int \limits_{\s_{re}}^1 \frac
{dg_{id}}{d\s} j_l (\bs) d\s.
 \label{eq:Gsid}
\end{equation}

To compute the coefficients $C_l$, we need to know the functions
\begin{equation*}
j_{l,id} = j_l + \cG_{l,id}, \quad j_{l,s} = j_l + \cG_{l,s}.
\end{equation*}
Consider first the function $j_{l,id}$. It is defined through the
derivative of function $g_{id} = 2b^{-1}\Pid/\Pz$, whose behavior
reflects the division of the interval between $\eta_{re}$ and
$\eta_0$ into two parts: it decreases from $2b^{-1}\o$ close to
$2b^{-1}\o_\infty$, then it slows down for a while, and then it
decreases again to the value $2b^{-1} \times 0.71$. Thus, its
derivative is negative and decreases in absolute value from a
finite value close to zero in the early period, to rise again to a
finite value in the late period. The behavior of the derivative of
$g_{id}$ is shown in the left panel of fig. 2, where it is
depicted by the heavy line denoted as `id'.
\begin{figure}[ht]
\centerline{\includegraphics[width=0.9\textwidth]{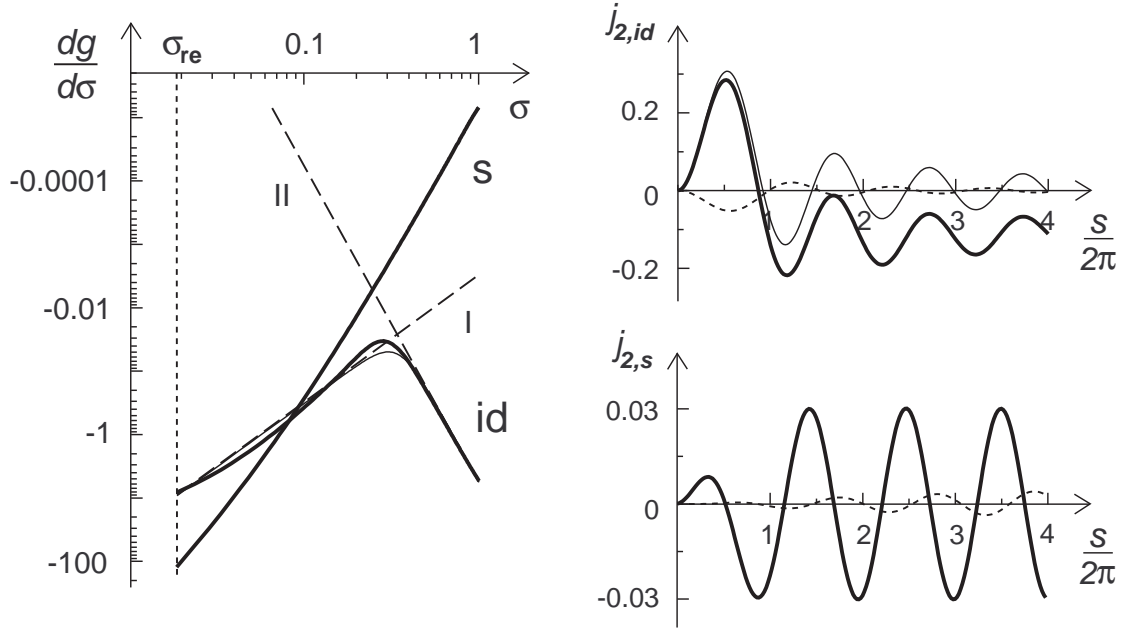}}
\centerline{\parbox {12.2cm}{\caption{\small Fluid and solid
contributions to the rate of suppression of superhorizon
perturbations after recombination (left) and the behavior of the
functions $j_{2,id}$ and $j_{2,s}$ describing how perturbations
with different wavenumbers contribute to the quadrupole (right)}}}
  \label{fig:j2}
\end{figure}
In the early and late periods the derivative can be approximated
by two power-like functions, one with negative power and another
one with positive power,
\begin{equation}
\Bl \frac {dg_{id}}{d\s} \Br_I = C_I \Bl \frac \s{\s_{re}}
\Br^{-2}, \quad \Bl \frac {dg_{id}}{d\s} \Br_\I = C_\I \s^5.
 \label{eq:gidap}
\end{equation}
The constants $C_I$ and $C_\I$ can be calculated by the method of
least squares and have values close to $dg_{id}/d\s$ at $\s =
\s_{re}$ and $\s = 1$ respectively. In the figure, the two
functions defined in equation (\ref{eq:gidap}) are depicted by the
dashed straight lines and their sum is depicted by the light line.

The behavior of the function $j_{2,id}$ is shown in the right
upper panel of fig. 2. The light line represents the function
$j_2$, the heavy line represents the function $j_{2,id}$ computed
with the help of the late-time contribution to $dg_{id}/d\s$, and
the dotted line represents the correction to the function
$j_{2,id}$ coming from the early-time contribution to
$dg_{id}/d\s$. The ratio $C_{2,id}/C_{2,id}^{SW}$, computed by
using the late-time contribution to $dg_{id}/d\s$ only, is 1.20,
and with the early-time contribution to $dg_{id}/d\s$ included
into the calculation, it drops to 0.91. The correct value is about
1.4. As seen from the figure, the value of
$C_{2,id}/C_{2,id}^{SW}$ could be raised either by suppressing the
early-time contribution to $dg_{id}/d\s$ or by enhancing the
late-time one. Our discussion of the evolution of perturbations in
the theory with decoupled radiation suggests that a certain
early-time suppression is necessarily produced by free-streaming.
A suppression at the level of 10 to 20 \%, which we have obtained
with the potentials $\Phi$ and $\Psi$ left out of the equation for
$\D T ({\bf x}, {\bf l}, \eta)$, does not help, but perhaps the
effect would become greater after including the potentials into
the calculation. Moreover, some late-time enhancement could arise
from the nonlinear integrated SW effect. To obtain results that
are at least qualitatively correct without complicating the theory
too much, we have mimicked the possible modifications of the
early-time and late-time contributions to $dg_{id}/d\s$ by
dropping the former contribution altogether and leaving the latter
contribution unchanged.

Let us now discuss the behavior of the function $j_{l,s}$. It is
defined through the function $dg_s/d\s$, depicted in the left
panel of fig. 2 by the heavy line denoted as `s'. Except for a
small region at the end point, the function $dg_s/d\s$ is
practically identical with the function $dg^\0_s/d\s$ computed
without cosmological constant. However, its value, and hence its
contribution to the function $j_{l,s}$, is negligibly small close
to the end point, therefore we can put it equal to $dg^\0_s/d\s$
everywhere. The value of $g_s^\0$ at $\s = \s_{re}$ is 1,
therefore we obtain by integration by parts that
\begin{equation*}
\cG_{l,s} = - j_l ((1 - \s_{re})s) + s \int \limits_{\s_{re}}^1
g_s^\0 j_l' (\bs) d\s.
\end{equation*}
The function $g_s^\0$ falls down rapidly with $\s$, for example,
it decreases from 1 to 0.1 at $\s = 3.6 \s_{re}$. As a result, the
integral in the expression for $\cG_{l,s}$ equals approximately
the product of the constant
\begin{equation*}
\int \limits_{\s_{re}}^1 g_s^\0 d\s \doteq \int
\limits_{\s_{re}}^\infty g_s^\0 d\s = \frac 35 \s_{re},
\end{equation*}
and the function $j_l'$ evaluated at the lower limit of the
integral. The resulting function $j_{l,s}$ is
\begin{equation}
j_{l,s} \doteq j_l (s) - j_l ((1 - \s_{re})s) + \frac 35 \s_{re} s
j_l' ((1 - \s_{re})s).
 \label{eq:jls}
\end{equation}
The approximate function $j_{2,s}$ is depicted in the right lower
panel of fig. 2 by the heavy line and the correction to it is
depicted in the same panel by the dotted line. The main
contribution of $j_{l,s}$ to the coefficient $C_l$ comes from the
interval $s \ll 1/\s_{re}$. For such $s$ we have $j_{l,s} \doteq
(8/5) \s_{re} s j_l'(s)$, with $s j_l'(s) = l j_l(s) = ls^l/(2l +
1)!!$ if $s \ll 1$ and $s j_l'(s) = s j_{l - 1} (s) = \cos (s -
\pi l/2)$ if $s \gg 1$. As a result, after completing a
semi-oscillation with small amplitude, the function $j_{l,s}$,
starts to oscillate uniformly with the amplitude $(8/5) \s_{re}$.

The ratio $C_l/C_{l,id}^{SW}$ as a function of $l$, computed for
extremal values of $\xs$ in the theory with cutoff at $\sm = 2$,
is depicted in fig. 3.
\begin{figure}[ht]
\centerline{\includegraphics[width=0.84\textwidth]{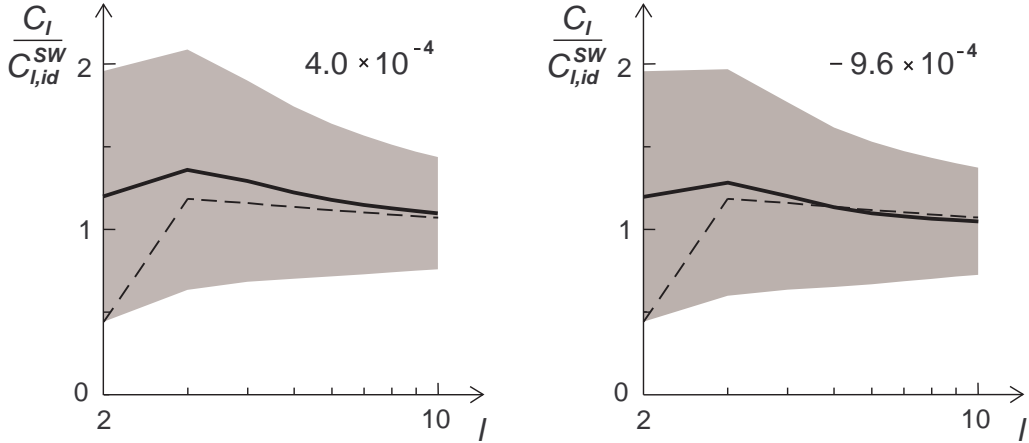}}
\centerline{\parbox {12.2cm}{\caption{\small Large-scale power
spectrum, computed for extremal values of shear modulus consistent
with the observed spectrum}}}
  \label{fig:aniz}
\end{figure}
The full and dashed lines represent theoretical and observational
values respectively, the shaded strips represent cosmic variance
and the numbers are the values of $\xi$. The main consequence of
including the integrated SW effect into the theory is that in the
formula for the coefficients $C_l$, the parameter $\xi_*$ becomes
multiplied by a function of order $\s_{re} = 1/\z_0$. As a result,
the interval of admissible values of $\xi_*$ is considerably wider
than in the theory without the integrated SW effect, with the
limits at $-145$ and 60. This is an enhancement approximately by
the factor $\z_0$; thus, due to the integrated SW effect the
maximum and minimum of $\xi$ become greater in absolute value
almost by two orders of magnitude.


\section{Conclusion}
 \label{sec:con}

We have calculated the large-ange part of the CMB power spectrum
in a universe containing radiation-like solid with constant shear
modulus to energy ratio $\xi$. For that purpose, we had to extend
the theory developed in \cite{bal} to the case when there is also
nonrelativistic matter and cosmological constant in the universe.
Taking into account ordinary Sachs-Wolfe effect only, we have
confirmed the conclusion of \cite{bal} that the parameter $\xi$
must satisfy $|\xi| \lesssim 10^{-5}$ to accommodate observations.
After including the integral Sachs-Wolfe effect into
considerations we have found that the constraint becomes relaxed
almost by two orders of magnitude. We have restricted ourselves to
large-angle anisotropies and did not investigate the position and
width of acoustic peaks. Obviously, for such small $|\xi|$ they
would look practically the same as in a universe filled with pure
fluid, unless the theory is modified, say, by endowing the solid
with a nonzero viscosity. The value of $|\xi|$ could be greater if
the effect of the solid was compensated by the tilt of the
primordial spectrum, but to assume such canceling of two
independent effects does not seem reasonable.

\end{document}